\documentclass{article}
\usepackage{amssymb}

\begin{document}
\author{Anatoly Konechny${\,}^{1}$ and  Albert Schwarz${\,}^{2}$\\
 \\
${}^{1}\,$Department of Physics, University of California Berkeley \\
and \\
Theoretical Physics Group, Mail Stop 50A-5101\\
LBNL, Berkeley, CA 94720 USA \\ 
konechny@thsrv.lbl.gov\\
\\
${}^{2}\,$Department of Mathematics, University of California Davis\\
Davis, CA 95616 USA\\
  schwarz@math.ucdavis.edu}

\title{\bf Compactification of  M(atrix) theory on noncommutative toroidal orbifolds }
\maketitle
\large
\begin{abstract}
\large
It was shown by A. Connes, M. Douglas and A. Schwarz that noncommutative tori arise 
naturally in consideration of toroidal compactifications of M(atrix) theory. 
A similar analysis of toroidal $\bf Z_{2}$ orbifolds leads to the algebra $B_{\theta}$ that 
can be defined as a crossed product of noncommutative torus and the group $\bf Z_{2}$ . 
Our paper is devoted to the study of projective modules over $B_{\theta}$ ( $\bf Z_{2}$-equivariant 
projective modules over a noncommutative torus). We analyze the Morita equivalence (duality) for 
$B_{\theta}$ algebras working out the two-dimensional case in detail. 
\end{abstract}
\large
\section{Introduction}

Our starting point is  BFSS (or IKKT) matrix model (\cite{IKKT}, \cite{BFSS}). 
To compactify this model on a $d$-dimensional torus $T^{d}$ in the bosonic sector of the theory 
one  should solve the following constraints 

\begin{equation} \label{eq1}
X_{j} + \delta_{ij}2\pi R_{j}\cdot {\bf 1} = U_{i}^{-1}X_{j}U_{i} 
\end{equation}
where $R_{1}, \dots , R_{d}$ are radii of $T^{d}$, $R_{i}=0$ for $i > d$. 
This equation does not have solutions in finite-dimensional matrices. Therefore,  
 we consider $X_{i}$ as Hermitian operators in Hilbert space $E$ and $U_{i}$ 
as unitary operators (see \cite{Taylor}). It was shown in \cite{CDS} that one can obtain an 
interesting class of solutions to (\ref{eq1}) where the operators $U_{\alpha}$ do not 
commute.  It is natural to assume that $U_{i}$ obey the relations 

\begin{equation}\label{ntorus}
U_{j}U_{k} = e^{2\pi i \theta_{jk}}U_{k}U_{j} \, .
\end{equation} 
Then the space $E$ along with the operators $U_{i}$ on it 
can be considered as a module over an algebra of (smooth) functions on a  noncommutative torus which we denote $A_{\theta}$; 
the operators $X_{i}$ specify a connection in this module (\cite{CDS}). We will restrict ourselves 
to projective modules (i.e. direct summands in finitely generated free modules). Then  one 
can thoroughly analyze compactifications of matrix models on noncommutative tori 
(see  \cite{CDS} - \cite{SeibWitt} and references therein). In the present paper we  will consider compactifications on
toroidal orbifolds with  an emphasis on orbifolds of the form $T^{d}/{\bf Z}_{2}$ where 
${\bf Z}_{2}$ acts  as a central symmetry $z \mapsto -z$. This symmetry should be 
implemented by a unitary (or antiunitary) operator $W$ in $E$ and the conditions (\ref{eq1}), 
(\ref{ntorus}) should be supplemented by  the relations
 
\begin{equation}\label{eq2}
-X_{i} = W^{-1} X_{i} W \, , \enspace i=1,\dots , d \quad X_{i} = W^{-1} X_{i} W \, , \enspace  i=d+1,\dots , 9 \, .
\end{equation}

A straightforward calculation shows that  the quantities $W^{2}$, $U_{i}WU_{i}W$, ($i=1, \dots, d$)  commute with all $X_{j}$'s.
 It is natural then to  set them to be constants
$$
W^{2} = c= \pm 1  \, , \quad 
U_{j}WU_{j}W = e^{2\pi i \lambda_{j}} \, .
$$
Redefining the fields $W \mapsto Wc^{-1/2}$, $U_{j} \mapsto U_{j} e^{-\pi i \lambda_{j}}c^{1/2}$ 
we get the following set of relations 
\begin{equation}\label{fullset}
U_{k}U_{j}U_{k}^{-1}U_{j}^{-1} = e^{2\pi i \theta_{kj}} \, , \quad WU_{i}W = U_{i}^{-1} \, , \quad W^{2} = 1 \, .
\end{equation}
Note that now the operator $W$ is chosen to be unitary. 
The conditions (\ref{fullset}) mean that the space $E$ can be considered as a module over 
a crossed product $B_{\theta}\equiv A_{\theta}\rtimes {\bf Z}_{2}$ (see Section 2 for a more thorough explanation). 
Formulas (\ref{eq1}), (\ref{eq2}) (for $i=1, \dots , d$) should be considered as a definition of a $B_{\theta}$-connection.

More generally one can consider 
a toroidal orbifold $T^{d}/G$ where $G$ is a finite group of linear transformations 
(we should assume that $T^{d} = R^{d}/L$ and $G$ transforms a lattice $L$ into itself). If the form $\theta$ is $G$ invariant
then  $G$ can be considered as an automorphism group of the noncommutative torus 
$T_{\theta}$ and it is possible to consider the crossed product of $T_{\theta}$ and 
$G$ (or twisted crossed product if we allow central extensions \cite{HoWu}, \cite{Doug_discr}). 
If $R_{ij}(g)$ are matrices representing the group elements  $g\in G$ on $L$, then in addition   to (\ref{eq1}) 
one has the following  bosonic and fermionic quotient conditions defining  the Matrix theory on $T^{d}/G$  
$$
R_{ij}(g) X_{j} = W^{-1}(g) X_{i} W(g)
$$
$$
 \Lambda_{\alpha \beta} (g) \psi_{\beta} = W^{-1}(g)\psi_{\alpha}W(g)
$$
where $\Lambda_{\alpha \beta} (g)$ is the matrix of spinor representation of $G$ satisfying 
$\Lambda^{\dagger} (g) \Gamma_{i} \Lambda (g) = R_{ij} (g) \Gamma_{j}$.  
In this paper our attention is concentrated on two-dimensional noncommutative toroidal $\bf Z_{2}$ orbifolds.
This is not a particularly 
 interesting 
example as far as physics is concerned, for one reason because  the  supersymmetry is broken completely.
However some of our results are valid (and are formulated) in any dimension. Other results in  dimensions 
$d>2$ can be deduced from $d=2$ results we obtained.

The $D0$ particles on the commutative orbifold $T^{4}/{\bf Z}_{2}$ were studied in papers \cite{RamgWald} and 
\cite{GLY}. 
In  the present paper we work in the framework of noncommutative geometry. 
However some of our results are new also in the commutative case.
We show how one can construct modules corresponding to a general action of 
the gauge group on Chan-Paton labels  describing various systems of branes wrapped on our noncommutative 
orbifold as well as branes ``sitting at the orbifold singularities'' (these words concerning interpretation 
should be understood in quotation marks as there are no points or geometric cycles in noncommutative geometry).

The paper is organized as follows. 
Section 2 contains a general description of modules over $B_{\theta}$ algebras. We discuss K-theory of 
 $B_{\theta}$ in section 3.
It was shown in \cite{ASMorita} that for the case of noncommutative tori 
the duality group is generated by the relation of Morita equivalence and coincides with the T-duality 
group $SO(d,d|{\bf Z})$. 
In section 4 we show how the constructions of  \cite{ASMorita} can be extended 
to  the algebras $B_{\theta}$. In sections 5 and 6 we turn to the detailed study of   the two dimensional case. 
In section 5 we calculate explicitly the $B_{\theta}$  K-theoretic invariants of  modules over two-dimensional orbifolds. 
We use these results to 
calculate the ($d=2$) duality group in section 6. The duality group we obtained is isomorphic to a semidirect product 
of the group $SO(2,2|{\bf Z})$ with a finite group. This finite group leaves the Chern numbers intact but 
acts on the other K-theoretic invariants in a nontrivial way. 
An explicit representation of the duality group in terms of generators and relations is given in the Appendix.


\section{Modules}

Let us consider  an algebra $A_{\theta}$ of smooth functions on a $d$-dimensional noncommutative torus. We  define   
$A_{\theta}$  as an associative algebra of formal expressions 
$$
\sum_{{\bf n}} C_{\bf n}U_{\bf n}
$$
 where $U_{\bf n}$ satisfy the relations

\begin{equation}\label{ntlattice}
U_{\bf n_{1}}U_{\bf n_{2}}= U_{\bf n_{1} + n_{2}}e^{\pi i <{\bf n_{1}} \theta {\bf n_{2}}>}
\end{equation}
and $C_{\bf n}$ tends to zero faster than any power of $|{\bf n}|$ as $\bf n$ tends to 
infinity. Here $\bf n$ runs over a lattice $D={\bf Z}^{d}$ which we assume to be embedded in ${\bf R}^{d}$ and $\theta$ is an antisymmetric 
$d\times d$  matrix. The algebra $A_{\theta}$ is equipped with an 
involution * transforming $U_{\bf n}$ to 
$U_{-{\bf n}}$. Note that for $\theta = 0$ this algebra  is isomorphic to the algebra of 
smooth 
functions on the standard (commutative) $d$-dimensional torus. One can say that we consider a 
``smooth part'' of a $C^{*}$-algebra $T_{\theta}$ with unitary generators $U_{\bf k}$, 
$k=1, \dots , d$ obeying $U_{k}U_{l}=e^{2\pi i \theta_{kl}}U_{l}U_{k}$.
A map $w:A_{\theta} \to A_{\theta}$ transforming $U_{\bf n}$ to $U_{-{\bf n}}$ can 
be extended to an 
automorphism of the algebra $A_{\theta}$ which we also denote by $w$. It is called a flip 
automorphism. 
The algebra $B_{\theta}$ is defined as a crossed product 
$A_{\theta}\rtimes_{w} {\rm Z}_{2}$. In other words $B_{\theta}$ is defined as an involutive algebra of 
formal linear combinations $a_{0} + a_{1}W$ where $a_{0}, a_{1}\in A_{\theta}$ and the operations 
with the symbol $W$ are governed by the relations $W^{*}=W$, $W^{2}= 1$, $WaW^{-1}=w(a)$. This 
means that the multiplication in $B_{\theta}$ is specified by the formula 

$$
(a_{0} + a_{1}W)(a_{0}' + a_{1}'W) =(a_{0}a_{0}' + a_{1}w(a_{1}')) + (a_{0}a_{1}'+ a_{1}w(a_{1}'))W 
\, .
$$

All algebras we are interested in are equipped with a trace coming from the standard trace  $\phi$ on 
a noncommutative torus: $\phi (\sum_{\bf n} C_{\bf n}U_{\bf n}) = C_{0}$. This trace which  
 we   denote $\tau$ is defined by the formula 
\begin{equation}\label{sttrace}
\tau(a_{0} + a_{1}W) = \phi (a_{0}) \, .
\end{equation}
One can prove that this is 
the only bounded trace on these algebras. It is important to notice however that there are new   
unbounded traces on the algebra  $B_{\theta}$. They can be defined by means of the 
following functionals  $\phi_{\bf \epsilon}\equiv \phi_{\epsilon_{1}, \dots, \epsilon_{d}}$ on 
$A_{\theta}$:
\begin{equation} \label{phi}
\phi_{\bf \epsilon}(U_{\bf n}) =\prod_{i=1}^{d}(-1)^{ \epsilon_{i}  n_{i}}  \, . 
\end{equation}
Here each $\epsilon_{i}$ takes values   0 or 1 that gives us $2^{d}$ functionals.. These 
functionals extended to $A_{\theta}$ by linearity satisfy the identity
$$
\tau_{\bf \epsilon}(ab)= \tau_{\bf \epsilon}(w(b)w(a)) \, .
$$ 
We obtain new $2^{d}$   unbounded traces $\tau_{\bf \epsilon}$ on $B_{\theta}$  by setting 
\begin{equation} \label{taueps}
 \tau_{\bf \epsilon}(a_{0} + a_{1}W) = 2\phi_{\bf \epsilon} (a_{1}) \, .
\end{equation} 
The factor of two here is inserted for normalization purposes that will be discussed in 
the next section. 
Note that in the commutative situation when $\theta = 0$ and $U_{\bf n} = e^{i \bf n \bf x}$ 
the functional $\phi_{\epsilon}$ is nothing but the functional evaluating the value at 
the fixed point $(\pi \epsilon_{1} , \dots , \pi \epsilon_{d})$. 

It is easy to see that a $B_{\theta}$-module can be described as an $A_{\theta}$-module $E$ with 
a unitary operator $W$ acting on it and satisfying   $W^{*}=W$, $W^{2}= 1$, $WaW^{-1}=w(a)$. 
Here we use the same notation $a$ for an element of $A_{\theta}$ and for the corresponding
operator  acting on $E$. 
It is well known (see  \cite{Julg}) 
that a $B_{\theta}$-module is projective if and 
only if the corresponding $A_{\theta}$-module is projective. Let us construct some examples of 
$B_{\theta}$-modules. Consider $A_{\theta}$ as a right $A_{\theta}$-module (one-dimensional free 
module). It can be equipped with an operator $W$ defined by the formula 
$W(a) = U_{\bf k}w(a)$ where ${\bf k}\in {\bf Z}^{d}$ is fixed. One can show that 
the modules corresponding  to vectors $\bf k$ and $\bf k'$ such that $k_{i}=k'_{i} \, ({\rm mod}\, 2)$ are isomorphic. 
Thus, without loss of generality we can assume that the coordinates $k_{i}$ take values $0,1$
$A_{\theta}$ equipped 
with the operator $W$ just defined can be considered as a $B_{\theta}$-module that will be denoted 
by $A_{\theta}^{\bf k}$. In the commutative $d=4$  case these modules were considered in \cite{RamgWald} 
 under the name ``representation II''.

Another  class of examples of $B_{\theta}$-modules comes from a generalization of 
the previous construction to the Heisenberg modules studied for the case of noncommutative tori 
in \cite{RieffelProj}. Let us remind the construction of Heisenberg modules. 
Let $G$ be a direct sum of ${\bf R}^{k}$ and an abelian finitely generated group, and let 
$G^{*}$ be  its dual group (the group of characters). In the most general situation 
$G={\bf R}^{p}\times {\bf Z}^{q}  \times F$ where $F$ is a finite group. Then 
$G^{*}\cong {\bf R}^{p}\times T^{q} \times F^{*}$. 
Consider a linear space $S(G)$ of functions on $G$ decreasing at infinity faster than any power. 
We define operators $U_{(\gamma, \tilde \gamma)}: S(G)\to S(G)$ labelled by a pair 
$(\gamma, \tilde \gamma)\in G\times G^{*}$ acting as follows 
\begin{equation}\label{U}
(U_{(\gamma, \tilde \gamma)}f)(x)=f(x+ \gamma )\tilde \gamma (x) \, . 
\end{equation}
One can check that the operators $U_{(\gamma, \tilde \gamma)}$ satisfy the commutation relations 
\begin{equation} \label{nt}
U_{(\gamma, \tilde \gamma)}U_{(\mu, \tilde \mu)}=   \tilde \mu (\gamma )\tilde \gamma^{-1} (\mu ) 
U_{(\mu, \tilde \mu)}U_{(\gamma, \tilde \gamma)}   \, .
\end{equation}
If  $(\gamma, \tilde \gamma)$ run over a $d$-dimensional lattice $\Gamma \subset G\times G^{*}$  (i.e. $\Gamma$ 
is a discrete subgroup such that $G\times G^{*}/\Gamma$ is compact)
then  formula (\ref{U}) defines a projective module over a $d$-dimensional noncommutative torus $A_{\theta}$ 
with 
\begin{equation}\label{cocycle}
exp(2\pi i \theta_{ij}) = \tilde \gamma_{i} (\gamma_{j} )\tilde \gamma_{j}^{-1} (\gamma_{i} )
\end{equation}
for a given basis $(\gamma_{i} , \tilde \gamma_{i})$ of the lattice $\Gamma$. 
Note that in the decomposition of $G$ above one has $2p + q = d$. 
 This $A_{\theta}$-module is called a Heisenberg module and denoted  $E_{\Gamma}$. 
It is more convenient to introduce  rescaled generators in   $E_{\Gamma}$ by the formula 
$\tilde U_{(\mu , \tilde \mu)} = U_{(\mu , \tilde \mu)} \tilde \mu (\frac{1}{2}\mu )$, $(\mu, \tilde \mu)\in \Gamma$. 
These generators satisfy in addition to (\ref{nt}) 
\begin{equation}\label{monomials}
\tilde U_{(\alpha , \tilde \alpha)}\tilde U_{(\beta , \tilde \beta)} = \tilde U_{(\alpha + \beta, \tilde \alpha \tilde \beta)} 
\tilde \beta \left( \frac{1}{2}\alpha \right) \tilde \alpha^{-1} \left( \frac{1}{2}\beta \right)  
\end{equation}
(cf. formula (\ref{ntlattice})).

Endomorphisms of a Heisenberg module $E_{\Gamma}$ have a basis consisting 
of operators $Z_{(\nu, \tilde \nu)}= U_{(-\nu, \tilde \nu^{-1})}$ with the pair
 $(\nu, \tilde \nu)$ belonging  to the dual lattice $\Gamma^{*}$, i.e. satisfying 
$\tilde \nu (\mu) = \tilde \mu (\nu)$ for any $(\mu,\tilde \mu)\in \Gamma$. 
Again, it is convenient to introduce rescaled operators $\tilde Z_{(\nu, \tilde \nu)} = \tilde U_{(-\nu, \tilde \nu^{-1})}$.

If one defines an operator $W_{0}:S(G) \to S(G)$ according to 
$(W_{0}f)(x)=f(-x)$, one obtains a $B_{\theta}$ module $E_{\Gamma , 0}$. One can also obtain 
new $B_{\theta}$-modules replacing   
the operator $W_{0}$ by an operator $W_{(\mu, \tilde \mu )}=\tilde Z_{(\mu, \tilde \mu )}W_{0}$, $(\mu, \tilde \mu )\in \Gamma^{*}$.
As one can easily check the Heisenberg module $E_{\Gamma}$ along with the operator  $W_{(\mu, \tilde \mu )}$ constitutes a
$B_{\theta}$-module  which we denote $E_{\Gamma , \mu, \tilde \mu}$.


\section{K-theory}

The $K$-groups of $B_{\theta}$ were calculated in \cite{FW}: 
$K_{0}(B_{\theta})\cong {\bf Z}^{3\cdot 2^{d-1}}$, $K_{1}(B_{\theta})=0$. 
One can check that the projective modules we constructed above generate the 
group $K_{0}(B_{\theta})$.
Let us show how to obtain a complete set of invariants of an element 
of $K_{0}(B_{\theta})$ (a complete set of topological numbers of a projective 
module over $B_{\theta}$).

Every  projective $B_{\theta}$-module $E$ can be represented as a direct summand of a free 
$B_{\theta}$-module $B_{\theta}^{n}$. In other words $E=eB_{\theta}^{n}$ where $e$ is a projection 
in $B_{\theta}^{n}$ (i.e. $e$ is an endomorphism such that $e^{*}=e$, $e^{2}=e$). 
The traces $ \tau_{\bf \epsilon}$ can be extended in a natural way to endomorphisms of 
$B_{\theta}^{n}$ (i.e. to $n\times n$ matrices with entries from $B_{\theta}$). It follows from 
well known results that the traces  $ \tau_{\bf \epsilon}(e)$ do not depend on the choice 
of projection $e$. More precisely $ \tau_{\bf \epsilon}(e)$ is an invariant of the $K$-theory 
class of the module $E$. Another invariant of the $K$-theory class of a $B_{\theta}$-module $E$ 
is the element $\mu(E)$ characterizing its class in the $K$-group 
$K_{0}(A_{\theta})=\Lambda^{even}({\bf Z}^{d})$ (see \cite{RieffelProj}). 
We will prove that the numbers $ \tau_{\bf \epsilon}(E)$ along with the element $\mu(E)$ 
characterize completely the $K$-theory class of the module $E$. 
Note that from the definition of the trace  (\ref{sttrace}) and the standard relation 
between $\mu(E)$ and the Chern character constructed with a help of $\tau$ it follows 
that the  representatives of $K_{0}(B_{\theta})$ corresponding to $\mu (E)$ are elements 
$1/2 \mu (E) \in 1/2 \Lambda^{even}({\bf Z}^{d})$ rather then $\mu$ itself. 
Formally this is  a question of normalization of the trace. We will stick to the chosen 
normalization which implies $1/2$ factors and denote the corresponding 
invariant by $\mu$. This normalization is consistent with 
the fractional branes picture ({\cite{frbr}, \cite{RamgWald}).

Let us calculate the invariants $ \tau_{\bf \epsilon}(E)$ in the case when 
$E=A_{\theta}^{\bf k}$. It is easy to check that 
$A_{\theta}^{\bf k}=e_{\bf k}B_{\theta}$ where $e_{\bf k}=\frac{1}{2}(1+ U_{\bf k}W )$. 
Thus, we obtain that 
\begin{equation} \label{newinv}
 \tau_{\bf \epsilon}(A_{\theta}^{\bf k})=
(-1)^{\epsilon_{1} k_{1}}\cdot \dots \cdot (-1)^{\epsilon_{d} k_{d}} \, .
\end{equation}
The other invariant of these modules  according to our normalization convention 
is $\mu = \tau (e_{\bf k}) = 1/2$.

Assigning to every projective module the numbers $ \tau_{\epsilon}(E)$ and the element 
$\mu(E)$ we obtain a homomorphism $T$ of $K_{0}(B_{\theta})$ into ${\bf R}^{3\cdot 2^{d-1}}$.
It follows from  formula (\ref{newinv}) and from the known facts about $K_{0}(A_{\theta})$ 
that the image of this homomorphism is a lattice in  ${\bf R}^{3\cdot 2^{d-1}}$. 
Combining this statement with the fact that $K_{0}(B_{\theta})\cong {\bf Z}^{3\cdot 2^{d-1}}$ 
we obtain that the homomorphism $T$ is injective.

For the case of the two-dimensional orbifold one can describe $K_{0}(B_{\theta})$ explicitly 
in terms of generators. For $d=2$ the invariant $\mu(E)$ is given by a pair of 
numbers $(p(E), q(E)) \in (1/2{\bf Z}, 1/2{\bf Z})$ related to the corresponding element in $K_{0}(A_{\theta})$ in an 
obvious way. We can define a mapping $T: K_{0}(B_{\theta}) \to {\bf R}^{6}$ by means of a vector function 
\begin{equation} \label{charges}
T(E) = (p(E), q(E); \tau_{00}(E), \tau_{01}(E), \tau_{10}(E), \tau_{11}(E))\, .
\end{equation}
It is proved by Walters (\cite{Walters}) that  for $d=2$ the image of $T$ is the additive subgroup of ${\bf R}^{6}$  
spanned by the vectors 

\begin{equation} \label{shortm} \
 (1/2, 0;1,1,1,1)\, , \enspace (1/2, 0;1,-1,1,-1)\, , \enspace (1/2,0;1,1,-1,-1)\, , \enspace  (1/2,0;1,-1,-1,1)
\end{equation}
\begin{equation} \label{otherm}
(1,0;0,0,0,0)  \, , \quad (0, 1/2;2,0,0,0) \, .
\end{equation}
Furthermore, Walters proved that for irrational $\theta$'s the positive cone $K_{0}^{+}(B_{\theta})$ of K-theory classes 
corresponding to modules is characterized by the property $dimE>0$.
It seems that  a similar result should be correct for $d>2$ and that one can prove it combining the results of the present paper 
with the arguments used in \cite{Walters}.
The generators in (\ref{shortm}) correspond to the modules $A_{\theta}^{\bf k}$. The first generator in 
(\ref{otherm}) is a free module and  the second generator  can be modeled on a Heisenberg 
module as we will show  in section 5.

Let us discuss K-theory of commutative orbifolds. 
 Consider a compact manifold $\cal M$ and a right action of a finite group $G$ on it.
The D-brane charges in IIA string theory  on $\cal M$ are associated with the elements of the K-group 
$K^{0}({\cal M})=K_{0}(C({\cal M}))$.
The D-brane charges of the orbifold theory obtained by means of factorization of $\cal M$ with 
respect to $G$ are related to $G$-equivariant K-theory on $\cal M$ (\cite{Witten}). 
For definiteness we will consider the complex K-theory. (The case of D-branes in type II theory 
is related to complex K-theory. In the type I string theory we obtain a real K-theory.)
The equivariant K-theory on $\cal M$ is equivalent to K-theory of noncommutative algebra 
$C({\cal M})\rtimes G$ (a noncommutative $C^{*}$ algebra defined as a crossed product of a 
commutative 
algebra $C({\cal M })$ of  functions on $\cal M$ and the group $G$ acting on this algebra).
The group $K(C({\cal M})\rtimes G)\otimes {\bf C}$ can be expressed in homological terms. 
Multiplication by $\bf C$ means that we disregard finite order elements in the K-group. 
For the standard K-theory there exists a map 
\begin{equation} \label{ch}
ch: K^{0}(X)\otimes_{\bf Z}{\bf Q} \to \prod_{j\in {\bf N}} H^{2j}(x; {\bf Q})
\end{equation}  
given by the Chern character. 
To define an analogue of this map in the equivariant case one should consider a kind of 
equivariant cohomology (``delocalized'' equivariant cohomology) $H^{i}_{G}({\cal M})$. 
Following \cite{BaumConnes} we will give a direct geometric definition of this 
cohomology. (It can also be defined  algebraically as cyclic cohomology of 
$C({\cal M})\rtimes G$.) The definition can be given in the following way. For $\gamma \in \Gamma$
we define $X^{\gamma}$ as the set of points $x\in X$ obeying $x \gamma = x$. Let us 
define $\hat X$ as a disjoint union $\hat X = \cup_{\gamma \in \Gamma} X^{\gamma}$. 
One can construct an action of the group $\Gamma$ on $\hat X$. (If $x\gamma = x$ then 
$(xg)(g^{-1}\gamma g)=xg$. Therefore, one can say that an element $g\in G$ specifies a map from $X^{\gamma}$ 
to $X^{g^{-1}\gamma g}$.) The group $H^{even}_{\Gamma}(X )$ can be defined as the $\Gamma$-invariant 
part of even-dimensional cohomology of $\hat X$ 
$$
H^{even}_{\Gamma}(X) = H^{even}(\hat X; {\bf C}) = \left(  \bigoplus_{\gamma \in \Gamma} 
H^{even}(X^{\gamma}; {\bf C})\right)^{\Gamma} \, .
$$ 
One can define the equivariant Chern character 
\begin{equation} \label{eqch}
ch_{\Gamma} : K^{0}_{\Gamma}(X) \to H^{even}_{\Gamma}(X)
\end{equation}
and prove that this character induces an isomorphism between $K^{0}_{\Gamma}(X)\otimes {\bf C}$ and
$H^{even}_{\Gamma}(X)$. The definition of $ch_{\Gamma}$ can be given in the following way. 
We represent an element of $K^{0}_{\Gamma}(X)$ as a $\Gamma$-equivariant vector bundle $E$ over $X$. 
This means that an element $g\in \Gamma$ acts on $E$ defining a linear map of a fiber over 
$x\in X$ into a fiber over $xg\in X$. In particular if $x\in X^{\gamma}$ then the fiber $E_{x}$ 
over $x$ is mapped into $E_{x}$ itself  by $\gamma$:
$$
\gamma : E_{x} \to E_{x} \, , \quad x\in X^{\gamma} \, .
$$
Let $\lambda_{1}, \lambda_{2}, \dots , \lambda_{s}$ be the set of distinct eigenvalues of this linear transformation.
Then, we have a direct sum  decomposition into eigenspaces: $E_{x} = E_{x}^{1}\oplus E_{x}^{2} \oplus \dots \oplus E^{s}_{x}$. 
At the level of vector bundles one has $E|_{X^{\gamma}} = E^{1}\oplus E^{2}\oplus \dots \oplus E^{s}$. 
The element $\gamma$ acts on $E^{i}$ by multiplication by $\lambda_{i}$. Following \cite{BaumConnes} we define 
$ch_{\Gamma}^{\gamma}(E)\in  H^{even}(X^{\gamma}; {\bf C})$ by 
\begin{equation} \label{char}
ch_{\Gamma}^{\gamma}(E)=\sum_{i=1}^{s}\lambda_{i} ch(E^{i})
\end{equation}
where $ch(E_{i})$ is the ordinary (non-equivariant) Chern character of $E^{i}$. 
Then the equivariant Chern character  (\ref{eqch}) is defined as the direct sum 
$$
ch_{\Gamma}(E)=\bigoplus_{\gamma \in \Gamma} ch_{\Gamma}^{\gamma}(E) \, . 
$$

Let us illustrate this construction in the case when 
$X=T^{d}$ and $G={\bf Z}_{2} = \{ \b{0},\b{1}\} $ where $\b{1}$ acts as $x\mapsto -x$. 
In this case $X^{\gamma}$ is equal to $T^{d}$ for $\gamma = \b{0}$ and $X^{\gamma}$ contains
$2^{d}$  points for $\gamma = \b{1}$. We see that $H^{even}(\hat X)$ consists of even-dimensional 
cohomology of $T^{d}$ and of $2^{d}$ summands $H^{0}(pt ; {\bf C}) = {\bf C}$. All elements 
of $H^{even}(\hat X)$ are ${\bf Z}_{2}$-invariant. Therefore, $H^{even}_{\Gamma}(T^{d} )$ has the dimension 
$2^{d-1} + 2^{d} = 3\cdot 2^{d-1}$.  One can prove that the K-group at hand has no torsion and 
therefore 
$$
K^{0}_{{\bf Z}_{2}}(T^{d}) = K_{0}(C(T^{d})\rtimes {\bf Z}_{2}) = {\bf Z}^{3\cdot 2^{d-1}} \, .
$$
This agrees with the K-group for noncommutative toroidal orbifold we considered. This is not 
surprising because the K-group being a discrete object usually does not change under continuous 
variations of  parameters. 
In all above considerations we talked about the group $K^{0}$. However one can prove  corresponding 
statements for the group $K^{1}$ and cohomology groups $H^{odd}$. In particular 
$H^{odd}_{{\bf Z}_{2}}(T^{d})=0$. This agrees with the fact that $K^{1}_{{\bf Z}_{2}}(T^{d})=0$.

One can relate the Chern character $ch_{\Gamma}$ in the case at hand with invariants 
$ \tau_{\epsilon}(E)$. This can be done in the following way. 
We defined above the $\tau_{\epsilon}(E)$-invariants in terms of a projector $e$ acting on a free module 
$B_{\theta}^{N}$. To realize the $B_{\theta}$-module $E$ as a $A_{\theta}$ module equipped with the 
action of $W$ we   map  $B_{\theta}^{n}$ to 
$A_{\theta}^{2N}$ by  
$(f_{i} + Wg_{i}) \mapsto (f_{i}, g_{i+N})$ , $i=1,\dots N$ , $f_{i}, g_{i} \in A_{\theta}$. 
The projector $e$ then is represented as a block $2n\times 2n$ matrix of the form
\begin{equation} \label{tildee}
\tilde e=\left(
\begin{array}{cc}
A&B\\
w(B)&w(A)
\end{array}
\right) 
\end{equation} 
and the operator $W$ acts on   $A_{\theta}^{2n}$ as 

\begin{equation} \label{descW}
W=\left(
\begin{array}{cc}
0&1\\
1&0
\end{array}
\right) \cdot w \, . 
\end{equation}

The projector $\tilde e$ commutes with $W$. Therefore, $W$ descends to an operator $\tilde W$ on the $A_{\theta}$-module 
$\tilde e A_{\theta}^{2n}$.   
 From (\ref{tildee}) and the definition of $\tau_{\epsilon}$ (\ref{taueps}) one immediately sees that $\tau_{\epsilon}(E) = 2tr \phi_{\epsilon} (B)$ 
where $tr$ denotes the matrix trace. 
Furthermore from (\ref{tildee}) and (\ref{descW}) we have 
$$
 \tilde e W = \left(
\begin{array}{cc}
B&A\\
w(A)&w(B)
\end{array}
\right) \cdot w \, .
$$
As was already noted, in the commutative case, the functional  $ \phi_{\epsilon}$  corresponds to
 evaluation at the fixed point $(\pi \epsilon_{1}, \dots , \pi \epsilon_{d})$. 
The operator $w$ acts on a 
function $f(x)$ on the torus by a reflection $f(x) \mapsto f(-x)$ leaving the values at fixed points unaltered. Hence, when 
restricting  the operator in the last formula to a fiber above a fixed point the operator $w$ can be dropped. Combining 
everything together we obtain    
 
\begin{equation} \label{commW}
\tau(\epsilon) (E) = 2tr \phi_{\epsilon}(B) =  
tr (\tilde e W) |_{(\pi \epsilon_{1}, \dots, \pi \epsilon_{d})} = tr \tilde W  |_{(\pi \epsilon_{1}, \dots, \pi \epsilon_{d})}
\end{equation}
Thus, indeed, in the commutative case  the invariants $\tau_{\epsilon}(E)$ correspond to characters  of the orbifold group 
representations on the fibers over fixed points in accordance with (\ref{char}). 
It follows then that these invariants are integer valued in the commutative case 
(note that the normalization factor of 2 in (\ref{taueps}) is necessary  for the integrality). For the noncommutative 
two-dimensional toroidal $\bf Z_{2}$ orbifold  it was proved in \cite{Walters} that the invariants $\tau_{\epsilon}(E)$ are 
integer valued. One can also prove the  integrality of $\tau_{\epsilon}(E)$ 
 in the noncommutative situation  in any dimension. Let us sketch the proof of this fact here. 
First, consider the case of rational $\theta$. Then the algebra $A_{\theta}$ is Morita equivalent to the algebra $C(T^{d})$ 
of functions on a commutative torus $T^{d}$. This means that there exists such a vector bundle over $T^{d}$ 
(a projective module over $C(T^{d})$) that $A_{\theta}$ can be considered as an algebra of endomorphisms of this bundle (module). 
The flip automorphism $w$ of the algebra $A_{\theta}$ can be realized as an involution of the bundle, i.e. as an involutive 
transformation of the total space of the bundle that covers the standard involution on $T^{d}$.
Using this realization of $A_{\theta}$ we can obtain a generalization of formula (\ref{commW}) to the case at hand.
This allows us to conclude that for rational $\theta$'s  as in the commutative case $\tau_{\epsilon}(E)$ is an 
integer. In order to analyze the case of irrational $\theta$'s we can use the fact that there exist families of 
projective modules continuously (even smoothly) depending on $\theta$. As we  just proved every module with a 
rational $\theta$ in such a family has integer values of $\tau_{\epsilon}$. By continuity this property extends to 
irrational $\theta$'s. There are sufficiently many families of this kind to prove that   $\tau_{\epsilon}$ is an integer 
for any projective module (we can use the fact that the K-group is spanned by Heisenberg modules that 
depend on $\theta$ in a continuous way in a proper range of $\theta$).


\section{Morita equivalence}
In this section we will show how Morita equivalences known for noncommutative tori can be extended to noncommutative 
 ${\bf Z}_{2}$ orbifolds.

Let $A$ and $\hat A$ be two involutive associative algebras. 
A $(A,\hat A)$-bimodule $P$ is said to  establish a   Morita equivalence between $A$ and 
$\hat A$. This means that the projective bimodule $P$ obeys the conditions 
\begin{equation} \label{P}
\bar P\otimes_{A}P\cong \hat A \, , \quad P\otimes_{\hat A}\bar P \cong A 
\end{equation}
where $\bar P$ is a $(\hat A, A)$-bimodule  that is  complex conjugated to $P$ that means that 
$\bar P$ consists of elements of $P$ and multiplications are defined as $\hat a (e) := (e) \hat a^{*}$, 
$ (e)a := a^{*} (e)$ where multiplications on the right hand sides are those defined for bimodule $P$.   
The algebras $A$ and $\hat A$ are said to be Morita equivalent if such $P$ satisfying  (\ref{P}) exists
The bimodule $P$ determines a one to one correspondence between $A$-modules and 
$\hat A$-modules by the rule 
$$
\hat E=E\otimes_{A}P \, , \quad E=\hat E\otimes_{\hat A}\bar P \, .
$$ 

For the case of noncommutative tori one can define a notion of complete Morita equivalence (\cite{ASMorita})
that allows one to transport connections between modules $E$ and $\hat E$. 
Let us remind here the basic  definitions. Let $\delta_{j}$, $j=1,\dots , d$ be a set of derivations of  $A_{\theta}$ 
specified by their action on generators  
$$
\delta_{j}U_{\bf n} = 2\pi i n_{j}U_{\bf n} \, .
$$ 
A connection on a projective module $E$ over $A_{\theta}$ can be defined as a set of operators 
$\nabla_{i}: E\to E$ satisfying a Leibniz rule 
$$
\nabla_{i}(ea) = (\nabla_{i}e)a + e(\delta_{i}a)
$$
for any $e\in E$ and any $a\in T_{\theta}$. 
We say that $(A_{\theta},A_{\hat \theta})$ Morita equivalence bimodule $P$ establishes a complete Morita equivalence if it
is endowed with operators $\nabla^{P}_{i}$ that determine a constant curvature connection 
simultaneously with respect to $A_{\theta}$ and $A_{\hat \theta}$, i.e. satisfy  

\begin{eqnarray}\label{biconnect}
&&\nabla^{P}_{i}(ae)=a\nabla^{P}_{i}e + (\delta_{i}a)e \, , \nonumber \\
&&\nabla^{P}_{i}(e\hat a)=(\nabla^{P}_{i}e)\hat a + e\hat \delta_{i} \hat a \, , \nonumber \\
&&[\nabla^{P}_{i},\nabla^{P}_{j}]=\sigma_{ij}\cdot {\bf 1} \, .
\end{eqnarray} 
Here  $\delta_{i}$ and $\hat \delta_{j}$ are standard derivations on $A_{\theta}$ and $A_{\hat \theta}$ respectively. 
Sometimes for brevity we will omit the word Morita in the term (complete) Morita equivalence bimodule.
If $P$ is a complete  $(A_{\theta},A_{\hat \theta})$ equivalence  bimodule then there exists a correspondence 
between connections on $E$ and connections on $\hat E$. An operator 
$\nabla_{i}\otimes 1 + 1\otimes \nabla_{i}^{P}$ on $E\otimes_{\rm C}P$ descends to a 
connection $\hat \nabla_{\alpha}$ on $\hat E = E\otimes_{A_{\theta}}P$. The curvatures of 
$\hat \nabla_{i}$ and $\nabla_{i}$ are connected by the formula 
$F_{ij}^{\hat \nabla}=\hat F_{ij}^{\nabla} + {\bf 1}\sigma_{ij}$.

Given an $(A_{\theta}, A_{\hat \theta})$ equivalence bimodule $P$ we say that it gives a
$(B_{\theta}, B_{\hat \theta})$ equivalence  if it is endowed with an involution 
$W^{P}$ satisfying
 
\begin{equation} \label{WP}
W^{P}aW^{P}=w(a) \, , \quad W^{P}bW^{P} = w(b)
\end{equation}
for any $a\in A_{\theta}$, $b\in A_{\hat \theta}$. 
Recall that we introduced  a notion of $B_{\theta}$ connection on a $B_{\theta}$-module $E$ in the following 
way.  Let $\nabla_{\alpha}$ be a connection on $E$ considered as an $A_{\theta}$-module. Then 
we call it a $B_{\theta}$ connection if it satisfies the additional condition 
$W\nabla_{\alpha}=-\nabla_{\alpha}W$. We say that $(A_{\theta},A_{\hat \theta})$ complete 
equivalence bimodule $P$ along with an operator $W^{P}$ satisfying (\ref{WP}) 
gives  a $(B_{\theta}, B_{\hat \theta})$ complete equivalence  
if  the 
connection $\nabla^{P}_{\alpha}$ satisfies
\begin{equation} \label{nablaP}
W^{P}\nabla^{P}_{\alpha}=-\nabla^{P}_{\alpha}W^{P}\, .
\end{equation} 
Given a triple $(P, \nabla^{P}_{\alpha}, W^{P})$ specifying a  $(B_{\theta}, B_{\hat \theta})$ complete equivalence  
it establishes  
 a correspondence between $B_{\theta}$ and $B_{\hat \theta}$ modules as well as 
between $B_{\theta}$ and $B_{\hat \theta}$ connections on them. Namely, if 
$\hat E= E\otimes_{A_{\theta}}P$, the involution $W$ in $E$ determines an involution 
$\hat W = W\otimes W^{P}$ in $\hat E$, i.e. a $B_{\theta}$-module $E$ generates a 
$B_{\hat \theta}$-module $\hat E$. It is straightforward to check that a $B_{\theta}$ connection 
in $E$ induces a $B_{\hat \theta}$ connection in $\hat E$.

If a triple $(P, \nabla_{\alpha}^{P}, W^{P})$ specifies a $(B_{\theta}, B_{ \theta'})$ complete 
equivalence and a triple  $( P',  \nabla_{\alpha}^{P'},  W^{P'})$ specifies a 
$(B_{ \theta'}, B_{\theta''})$ complete equivalence, then the tensor product 
$ P'' = P\otimes_{B_{\theta'}}  P'$ along with the appropriate connection $ \nabla_{\alpha}^{P''}$ and 
involution $  W^{P''}$ determines $(B_{\theta}, B_{ \theta''})$ complete equivalence. This means that we 
can consider a groupoid of equivalences.

Notice that a  pair $(P,W^{P})$ that gives a  $(B_{\theta}, B_{\hat \theta})$ equivalence 
is not a Morita equivalence $(B_{\theta}, B_{\hat \theta})$-bimodule in the sense of the 
general definition (it is not even a $(B_{\theta}, B_{\hat \theta})$-bimodule). However 
one can construct a $(B_{\theta}, B_{\hat \theta})$ Morita equivalence bimodule $\tilde P$ 
in the following way. As a vector space $\tilde P$ is a direct sum of two copies of $P$ and 
a $(B_{\theta}, B_{\hat \theta})$-bimodule structure is defined by the formulas
$$
(p_{1},p_{2}) b = (p_{1}b, p_{2}b)\, , \quad a (p_{1},p_{2})=(ap_{1},w(a)p_{2}) \, ,
$$
$$
(p_{1},p_{2})\hat W = (W^{P}p_{2}, W^{P}p_{1})\, , \quad W(p_{1},p_{2}) = (p_{2},p_{1})
$$
where $a\in A_{\theta}$, $b\in A_{\hat \theta}$. 
It is clear from the above consideration that every basic Heisenberg module gives a 
  $(B_{\theta}, B_{\hat \theta})$ complete equivalence. It has a unique (up to an overall sign) operator $W^{P}$ 
given by the formula 
$$
W^{P} f(x) = f(-x) \, , \qquad f(x) \in L_{2}(G) \, , 
$$
and the standard choice of constant curvature connection (\cite{RieffelProj}) $\nabla_{\alpha}$ satisfies (\ref{nablaP}). 


\section{Modules in the two-dimensional case}
In this section we will calculate the invariants $\tau_{\epsilon}(E)$ for a certain class 
of modules over a two-dimensional toroidal $\bf Z_{2}$-orbifold. 
For the purposes of calculating the duality group it suffices to have information about 
the modules $A_{\theta}^{\bf k}$ and $E_{\Gamma , \mu \tilde \mu}$ defined in Section 2.
The invariants $\tau_{\epsilon}(A_{\theta}^{\bf k})$ are given by formula (\ref{newinv}). 
In this section we will calculate the invariants of  the modules $E_{\Gamma , \mu \tilde \mu}$.

In two dimensions the K-theory class of a module $E$ over  $A^{2}_{\theta}$ is labeled by two topological 
numbers $(n, m)$ such that $dimE= n - m\theta > 0$. A module with the numbers $n$ and $m$ relatively prime 
is  called basic (see \cite{KS}, \cite{KS3}). For irrational $\theta$  any  module can be represented 
as a direct sum of isomorphic basic modules.
We define a class of $B_{\theta}$-modules with an underlying 
basic $A_{\theta}$-module  by the following explicit formulas defining the action of the generators 
on $L_{2}({\bf R}\times {\bf Z}_{m})$
 
\begin{eqnarray}\label{U's}
(U_{1}\phi)_{j}(x) &=& \phi_{j-1}\left( x-\frac{n}{m} + \theta \right) \nonumber \\
(U_{2}\phi)_{j}(x) &=& \phi_{j}(x) e^{2\pi i(x-jn/m)} 
\end{eqnarray}

\begin{equation} \label{W}
W \phi_{j} (x) = exp\left( \pi i k_{1}\frac{a\theta - b}{n-m\theta}k_{2}\right) Z_{1}^{k_{1}}Z_{2}^{k_{2}} \phi_{-j}(-x) \, .
\end{equation}
Here $j=1, \dots , m$ is a ${\bf Z}_{m}$-index and should be understood as a $mod(m)$ equivalence class, 
$a$ and $b$ are integers such that $an-bm=1$ and
the operators $Z_{1}$, $Z_{2}$ are generators of endomorphisms  defined by the formulas 

\begin{eqnarray} \label{Z's}
(Z_{1}\phi)_{j}(x)& =& \phi_{j-a}\left( x-\frac{1}{m} \right) \nonumber \\
(Z_{2}\phi)_{j}(x)&=& \phi_{j}(x) exp( 2\pi i \left( \frac{x}{n-m\theta } - \frac{j}{m}\right) ) \, .
\end{eqnarray} 
The modules  specified by (\ref{U's}), (\ref{W}), (\ref{Z's}) are a particular type of the modules 
$E_{\Gamma, \mu, \tilde \mu}$ introduced above.
 Note that the components of the vector $\bf k$ matter only modulo 2 in the sense that the corresponding 
modules are isomorphic.

We would like to calculate now the invariants $\tau_{\epsilon}$ of these modules. As it was proved by 
Walters (\cite{Walters}) for the two-dimensional case  and as we argued above in  general  
the invariants $\tau_{\epsilon}(E)$ 
are integer valued. On the other hand the construction (\ref{U's}), (\ref{W}), (\ref{Z's}) of the module  
depends on $\theta$ in a continuous way provided $n-m\theta \ne 0$. Thus, we see that 
at least for the modules with $n>0$ we can reduce the calculation of 
$E_{n,m, {\bf k}}$ to the commutative case when $\theta = 0$. In the commutative situation as it was explained 
above   $\tau_{\epsilon}(E)$ coincides with the trace of $W$ evaluated at a fixed point. 
To calculate this trace we  find  a set of common eigenvectors of $U_{1}$ and $U_{2}$, restrict $W$ to 
a subspace of eigenvectors corresponding to each  fixed point and calculate its trace.
The result is given by the formula 
\begin{eqnarray}\label{trW}
&&\tau_{\epsilon}(E)=trW|_{\epsilon_{1}, \epsilon_{2}} =
 \frac{1}{2}(1 - (-1)^{n})(-1)^{k_{1}\epsilon_{1} + k_{2}\epsilon_{2}} 
(-1)^{bk_{1}k_{2} + m\epsilon_{1}\epsilon_{2}} + \nonumber \\
&& + (1 + (-1)^{n})(-1)^{ak_{1}k_{2}}\delta_{\bar k_{2}, \epsilon_{1}}
\delta_{\bar k_{1}, \epsilon_{2}} 
\end{eqnarray}
where $\bar k_{1}$, $\bar k_{2}$ stand for the $mod (2)$ classes.
We must note here that the expression (\ref{trW}) depends on the choice of integers $a$ and $b$ 
through the overall sign. This happens because the generator $W$ given in (\ref{W}) is fixed 
up to a sign. Note that this sign factor is sensitive to the parities of $a$ and $b$ only. 
To fix  the sign of $W$ it is convenient to make the following choice of the parities of $a$ and $b$ 
depending on the parities of $n$ and $m$. We set $a$ even and  $b$ odd if $n$ and $m$ are both odd 
and $a$, $b$ both odd otherwise. We denote these modules  $E_{n,m, {\bf k}}$. With this convention 
we have 
\begin{eqnarray}\label{trW'}
\tau_{\epsilon}(E_{n,m, {\bf k}}) &=&
 \frac{1}{2}(1 - (-1)^{n})(-1)^{k_{1}\epsilon_{1} + k_{2}\epsilon_{2}} 
(-1)^{k_{1}k_{2} + m\epsilon_{1}\epsilon_{2}} + \nonumber \\ 
 && + (1 + (-1)^{n})(-1)^{k_{1}k_{2}}\delta_{\bar k_{2}, \epsilon_{1}} 
\delta_{\bar k_{1}, \epsilon_{2}} \, .
\end{eqnarray}

So far the above arguments prove the expression (\ref{trW}) to be applicable only for the case when 
$m$ and $n$ are relatively prime and $n>0$. If the second condition does not hold then the module 
becomes virtual for $\theta =0$. However we claim that the formula (\ref{trW'})  also holds in those cases. 
To perform a calculation we need to deform our module  to a commutative one. We do not necessarily have 
to go to $\theta=0$ but to some other integer. We can choose $\theta^{*} \in {\bf Z}$ in such a way that 
$dim E = n- m\theta^{*}>0$ and then perform a calculation. We have to be careful though because modules 
over tori with $\theta$ and $\theta + l$, $l\in {\bf Z}$ are identified in a nontrivial way.     
Namely a module with Chern numbers $(n, m)$ over $A_{\theta}$ corresponds  to   a module characterized by 
$(n+ m, m)$ over $A_{\theta + 1}$. It is not hard to find the identification of invariants $\tau_{\epsilon}$ 
that corresponds to the shift $\theta \mapsto \theta + 1$. This shift induces an outer automorphism 
of  $A_{\theta}$ specified by $U_{\bf n} \mapsto (-1)^{n_{1}n_{2}}U_{\bf n}$. It is a matter of a simple calculation 
to find what happens to the functionals (\ref{phi}) under this automorphism. They get transformed by a  linear 
transformation
\begin{equation} \label{transf}
\phi'_{\epsilon} = \frac{1}{2} \sum_{\eta_{1}, \eta_{2}=0,1} (-1)^{(\epsilon_{1} + \eta_{1})(\epsilon_{2} + \eta_{2})} 
\phi_{\eta} \, .
\end{equation} 
It follows then from (\ref{taueps}) and the fact that invariants $\tau_{\epsilon}(E)$ are traces of the corresponding 
projectors that $\tau_{\epsilon}(E)$ transform by the same linear transformation. 
As an illustration let us show how  one can calculate the invariants $\tau_{\epsilon}(E)$ for  the module 
$E=E_{n,m, {\bf k}}$ with $n=0$, $m=1$, ${\bf k} = 0$. Such a module does exist for $\theta <0$. 
We can first go to $\theta = -1$ and use formula (\ref{trW'}) for $n=1$, $m=1$ and then apply transformation 
(\ref{transf}) to the result. The calculation yields $( \tau_{\epsilon}(E_{0,1,0}) ) = (2, 0, 0 ,0)$.
 The K-theory class of $E_{0,1,0}$ is characterized by $T(E_{0,1,0})=(0,1/2;2,0,0,0)$ and thus it represents a K-theory generator 
defined in (\ref{otherm}). In general one can show that formula (\ref{trW'}) is applicable to any basic module over any $A_{\theta}$.

The calculation leading to formula (\ref{trW}) goes as follows.
Although we are interested in this section in two dimensional tori only, for the future purposes we will 
give a formula for common eigenvectors of $U_{i}$ that is valid for an arbitrary Heisenberg module over (commutative) $T^{d}$.  
Let $E$ be a Heisenberg module defined with a help of a lattice $\Gamma \in G\times G^{*}$. 
An eigenvector $\phi(x)$ (that does not lie in the $L_{2}$ space, but is a generalized function) 
satisfies 
$$
\tilde U_{\mu, \tilde \mu} \phi = C (\mu , \tilde \mu ) \phi \, . 
$$
As the operators $\tilde U_{\mu, \tilde \mu}$ form a commutative group we can construct a common eigenvector out of any 
function $\phi_{0}(x)$ by the following averaging procedure

\begin{equation} \label{eigenvect}
\phi(x) = \sum_{(\mu, \tilde \mu )\in \Gamma} (\tilde U_{(\mu, \tilde \mu )}\phi_{0}) (x)  \lambda(-\mu, \tilde \mu^{-1}) 
\end{equation}  
where $\lambda$ is an arbitrary element of the dual group $(G\times G^{*})^{*}$. As one readily checks 
the corresponding eigenvalues are $C (\mu , \tilde \mu ) = \lambda(\mu, \tilde \mu)$. 
The dual group $(G\times G^{*})^{*}$ can be identified with the group $G\times \tilde G$ itself. Thus, every 
character $\lambda$ is of the form
 
\begin{equation}\label{lambda}
\lambda (\mu , \tilde \mu ) = \tilde \xi (\mu ) \tilde \mu^{-1} (\xi )
\end{equation}
where $(\xi , \tilde \xi) \in G\times G^{*}$. Evidently the eigenvalues $C (\mu , \tilde \mu )$ 
 are in a  one-to-one correspondence with 
the points in $G\times G^{*}/\Gamma^{*}\cong T^{d}$ as of course one should expect.
Any function $\phi_{0}$ substituted into (\ref{eigenvect}) gives a common eigenvector but most of the eigenvectors corresponding to 
different $\phi_{0}$'s are linearly dependent. It is convenient to choose a set of $\phi_{0}$'s 
in the form $\phi_{0}^{(i)}(x) = \tilde Z_{(\nu_{i}, \tilde \nu_{i})} \delta (x)$ where $\delta(x)$ is a delta function and 
$(\nu_{i}, \tilde \nu_{i})$ corresponds to a choice of representative in $\Gamma^{*}/\Gamma$. 
Since the operators $\tilde Z_{(\nu, \tilde \nu)}$  
commute with all $\tilde U_{(\mu, \tilde \mu)}$'s the index $i$ corresponds to vectors in the same 
eigenvalue subspace. However it  still overcounts the solutions.  
One can identify the correct degeneracy label when working with explicit constructions of modules.

Applying the formulas (\ref{eigenvect}), (\ref{lambda}) to the modules defined by (\ref{U's}) for $\theta=0$
we obtain the following expressions for the eigenvectors 

\begin{equation}\label{2Deigenvect}
\phi_{j}^{\sigma , r}(x) = \sum_{s\in {\bf Z}} \delta \left(x- \frac{n}{m}j-r-ns- \frac{\sigma_{2}}{2\pi} \right) e^{-im/n\sigma_{1}x}
\end{equation}
where $r=1,\dots , n$ labels the degeneracy, $0\le \sigma_{i} < 2\pi$.
The corresponding eigenvalues of $U_{1}$ and $U_{2}$ are $e^{i\sigma_{1}}$ and $e^{i\sigma_{2}}$ respectively.
For any action of $W$ on this module the subspace of eigenfunctions corresponding to a fixed point 
$\sigma_{i} = \pi \epsilon_{i}$, $\epsilon_{i}=0,1$  is invariant. One can calculate now the trace of 
W defined by  (\ref{W}) (with $\theta=0$) restricted to   these subspaces. The result is  given by formula (\ref{trW}).


\section{Duality group}
In this section we consider the $d=2$ case.
As we have seen in section 4 $(B_{\theta}, B_{\hat \theta})$ equivalence is generated by a 
 $(A_{\theta}, A_{\hat \theta})$ equivalence bimodule equipped with an involution $W^{P}$ obeying 
(\ref{WP}). More precisely every bimodule of such a kind determines an isomorphism of K-groups 
$K(B_{\theta})$ and $K(B_{\hat \theta})$. We will construct a group that governs these isomorphisms 
(the duality group). Both of the K-groups $K(B_{\theta})$ and $K(B_{\hat \theta})$ in the $d=2$ case 
are isomorphic to ${\bf Z}^{6}$. Their elements can be characterized by integers 
$(n,m, \tau_{00}(E), \tau_{01}(E), \tau_{10}(E), \tau_{11}(E))$ where $n=2p(E)$, $n=2q(E)$ 
(see (\ref{charges})).

Let us   formulate the answer we obtained. 
The whole duality group $G$ is generated by two subgroups each isomorphic to $SL(2, {\bf Z})$;  we 
denote them by $SL(2,{\bf Z})_{M}$ and $SL(2, {\bf Z})_{T}$. 
The first subgroup is just the modular group of the torus. In two dimensions it  leaves 
the parameter $\theta$ and the invariants $(n,m)$ unchanged but acts on  $\tau_{\epsilon}(E)$ 
in the following way 
\begin{equation} \label{modular}
( A \tau )_{\epsilon} = \tau_{\bar A^{t} \epsilon} \, .
\end{equation}
Here $A$ is a matrix from $SL(2,{\bf Z})_{M}$ and $\bar A$ stands for the matrix with entries reduced modulo 2, 
such a matrix acts naturally on the vector space $({\bf Z}_{2})^{2}$   which the label $\bf \epsilon$ is an element of.

The second subgroup   $SL(2, {\bf Z})_{T}$ is generated by two transformations $\tau$ and $\sigma$ that 
act on the torus parameter $\theta$ as 
\begin{equation}
\tau : \theta \mapsto \theta + 1 \, , \qquad \sigma : \theta \mapsto -\frac{1}{\theta} \, . 
\end{equation}
The action of these generators on the topological numbers reads as 
\begin{equation} \label{nmtransf}
\tau : (n, m) \mapsto (n-m, m) \, \qquad \sigma : (n, m) \mapsto (-m, n) \, ,
\end{equation}

\begin{equation} \label{tauontraces}
\tau : \tau_{\epsilon} \mapsto \frac{1}{2}\sum_{\eta_{1}, \eta_{2} = 0,1 } 
(-1)^{(\epsilon_{1} + \eta_{1})(\epsilon_{2} + \eta_{2})} \tau_{\eta_{1}, \eta_{2}} \, ,
\end{equation}
\begin{equation} \label{sigma}
\sigma :  \tau_{\epsilon} \mapsto \frac{1}{2}\sum_{\eta_{1}, \eta_{2} = 0,1 } 
(-1)^{\epsilon_{1}\eta_{2} + \eta_{1}\epsilon_{2}} \tau_{\eta_{1}, \eta_{2}} \, . 
\end{equation}

The whole duality group being restricted to the invariants $(n,m)$ only, gives the 
familiar spinor  representation of the group $SO(2,2|{\bf Z})$ isomorphic to 
the direct product of $SL(2, {\bf Z}) \times SL(2, {\bf Z})$ where the components 
correspond to  the restrictions of  the subgroups $SL(2,{\bf Z})_{M}$ and $SL(2, {\bf Z})_{T}$.
 In the whole duality group the elements of these subgroups do not commute in general.
However one can find a new set of generators that generates a subgroup isomorphic to 
$SO(2,2|{\bf Z})$. The duality group then is isomorphic to a semidirect product 
$SO(2,2|{\bf Z})\ltimes H$ where H is a finite group  
that  can be characterized as a central extension of the group $({\bf Z}_{2})^{4}$. 
In other words there exists an exact triple
$$
0\to {\bf Z}_{2} \to H \to ({\bf Z}_{2})^{4} \to 0 \, .
$$
We relegate the construction of this $SO(2,2|{\bf Z})$ subgroup and subgroup $H$ to the Appendix.

We now proceed to the description of Morita equivalences  and to computations. 
First, there are $(A_{\theta}, A_{\theta})$ bimodules corresponding to the modular transformations of a torus. 
They are constructed in the following way.  To avoid confusion 
 we will denote by $U^{l}_{\bf n}$ the left action of the first torus generators  and 
by $U^{r}_{\bf n}$ the action of those of the second one acting on the right.  Let $R\in SL(2, {\bf Z})$. 
The bimodule $E_{R, \theta}$ consists of elements $a\in A_{\theta}$ 
and the action of the tori is defined as follows 

\begin{equation} \label{Rbimod}
U^{l}_{\bf n} (a) = U_{\bf n} a \, , \qquad (a)U_{\bf n}^{r} = a U_{R({\bf n})} \, .
\end{equation}   
With this definition the product $E_{R_{1}, \theta} \otimes_{A_{\theta}} E_{R_{2}, \theta}$
is isomorphic to $E_{R_{1}R_{2}, \theta}$; a canonical isomorphism is specified by $a\otimes 1 \mapsto a$. 
The bimodule $E_{A, \theta}$ can be equipped with an operator $W^{P}$ satisfying (\ref{WP}) and the connection $\nabla^{P}$ 
satisfying (\ref{biconnect}), (\ref{nablaP}) by the formulae
 
\begin{equation} \label{w0}
W^{P} (a) = w(a) \, , \quad \nabla^{P}_{i} (a) = \delta_{i}(a) 
\end{equation}
where $\nabla_{i}$ is a standard differentiation on $A_{\theta}$ defined by 
$\delta_{j}U_{\bf n} = 2\pi i n_{j} U_{\bf n}$.
It is pretty straightforward now to calculate from (\ref{Rbimod}) the transformation of the functional $\phi_{\epsilon}$ 
defined by (\ref{phi}). It is easy to see from the structure of the bimodule that in the case at hand 
the transformation law for the traces $\tau_{\epsilon}$ is the same as the one for $\phi_{\epsilon}$ and is given 
by  (\ref{modular}). 
Note that in two dimensions modular transformations do not change the parameter $\theta$. But in 
$d$ dimensions the matrix $\theta$ transforms as $\theta \mapsto R^{t}\theta R$, $R\in SL(d, {\bf Z})$ 
and the described above bimodules should be considered as  $(A_{\theta}, A_{R^{t}\theta R})$-bimodules.

The $(A_{\theta}, A_{\theta + 1})$- bimodules corresponding to transformation $\tau$ defined above 
consist of elements $a\in A_{\theta}$ and the tori action reads 
\begin{equation} \label{taubimod}
U^{l}_{\bf n} (a) = U_{\bf n} a \, , \qquad (a)U_{\bf n}^{r} = a U_{\bf n}(-1)^{n_{1}n_{2}} \, .
\end{equation}
The operator $W^{P}$ and connection $\nabla^{P}$ are the same as the ones defined above by  formula (\ref{w0}).
This Morita equivalence corresponds to an outer automorphism that was already discussed in the previous section.   
 The corresponding transformation of the traces follows from the transformation (\ref{transf}) 
of the functionals $\phi_{\epsilon}$ induced by this automorphism. As a result we get formula (\ref{tauontraces}).

Finally the equivalence bimodule realizing the transformation $\sigma$ is constructed as follows. 
The $(A_{\theta}, A_{-1/\theta})$-equivalence bimodule $E_{\sigma, \theta}$ is a Heisenberg bimodule defined on $L_{2}({\bf R})$ by 
the formulae 
\begin{eqnarray}\label{UZ}
(U_{1}^{l}\phi)(x) &=& \phi( x + \theta) \, ,\nonumber \\
(U_{2}^{l}\phi)(x) &=& \phi(x) e^{2\pi ix} \, ,\nonumber \\
(U_{1}^{r}\phi)(x)& =& \phi(x)e^{2\pi i x/\theta} \, ,\nonumber \\
(U_{2}^{r}\phi)(x)&=& \phi(x-1) \, .
\end{eqnarray} 
The above form of the bimodule can be deduced from the constructions of \cite{ASRieffel}.
The operator $W^{P}$ acts as $W^{P}\phi (x) = \phi(-x)$ and the connection $\nabla^{P}$ is the standard constant 
curvature connection constructed out of $\frac{\partial}{\partial x}$ and $x$ operators (\cite{RieffelProj}, \cite{CDS}). 
In order to calculate the action of $\sigma$ on $\tau_{\epsilon}$ one   can first  tensor this bimodule with any one of 
$A_{\theta}^{\bf k}$ modules. The result is the right Heisenberg 
module defined in (\ref{UZ}) equipped with an operator $\hat W$ given by a formula analogous to (\ref{W}). 
More precisely this module is 
isomorphic to the module $E_{0,1,{\bf k}}$ introduced above. 
Let us denote this module $E_{\bf k}=A_{\theta}^{\bf k}\otimes_{A_{\theta}}E_{\sigma, \theta}$. 
Thus, using (\ref{newinv}) and (\ref{trW}) we can calculate the transformation of the invariants $\tau_{\epsilon}(E)$ in this case. 
Furthermore one can use the fact that $\sigma^{2}=1$ and thus calculate what happens with $\tau_{\epsilon}$ when one applies $\sigma$ 
to  $E_{\bf k}\cong E_{0,1,{\bf k}}$. 
As it was explained in sections 3, 5 the modules $A_{\theta}^{\bf k}$, the rank one  free module and the Heisenberg module $E_{0,1,{\bf k}}$
 generate the whole K-theory group. So the information we have suffices to deduce the transformation of 
$\tau(E)$ for any module $E$. The result is given by (\ref{sigma}). 
This concludes the derivation of the duality group transformations 
corresponding to elements of $SL(2,{\bf Z})_{M}$ and $SL(2, {\bf Z})_{T}$ subgroups. 
Let us comment briefly on the form of the transformation ({\ref{sigma}). One can introduce a discrete (${\bf Z}_{2}\times {\bf Z}_{2}$) 
Fourier transform
$$
\hat f({\bf \epsilon}) = \frac{1}{2}\sum_{\nu \in ({\bf Z}_{2})^{2}} e^{i\pi(\nu_{1}\epsilon_{1} + \nu_{2}\epsilon_{2})}f({\bf \nu})
$$
defined on functions on ${\bf Z}_{2}\times {\bf Z}_{2}$.
Then $\sigma$ is just a combination of such a ${\bf Z}_{2}\times {\bf Z}_{2}$ Fourier transform and a rotation by $\pi$ that 
exchanges $\epsilon_{1}$ with $\epsilon_{2}$.

\begin{center} {\bf Acknowledgements} \end{center}

We are indebted to C.~Farsi, M.~Rieffel and K.~Wendland  for useful discussions.  

\section*{Appendix. Duality group in terms of generators and relations}
We proved in the main text of the paper that the whole duality group is generated by transformations specified by 
(\ref{modular}) - (\ref{sigma}).
 One may  not be completely satisfied with this description of the group. In this appendix we introduce more generators 
(that can be obtained from commutators of the ones introduced above) and give a set of relations characterizing the group.

In addition to the equivalence bimodules we described in the last section of the paper 
there are also pairs $(P, W^{P})$ corresponding to automorphisms of $B_{\theta}$ algebras that leave the parameter 
$\theta$ unchanged. However, when applied to modules this transformations relabel the invariants $\tau_{\epsilon}(E)$.
Moreover the duality transformations corresponding to these automorphisms occur when one commutes the transformations 
from the subgroups  $SL(2, {\bf Z})_{M}$ and  $SL(2, {\bf Z})_{T}$. 
Let us make here the following  remark concerning ideology. In general if we consider an associative algebra $A$ 
any automorphism of $A$ generates a Morita equivalence bimodule. The internal automorphisms give rise to trivial equivalences. 
The external ones may result in a nontrivial relabeling of the elements in $K_{0}(A)$ and should be included. 
In the case at hand as we are not trying to study all possible Morita equivalences but rather extending the once known for tori, 
we consider only those additional bimodules that are necessary for the group closure. 
For completeness let us give here a construction of these additional bimodules. 
There are two types of them. The first type comes from automorphisms $\eta_{i}: U_{\bf n} \mapsto (-1)^{n_{i}}U_{\bf n}$. 
The bimodule $E_{\eta_{i}, \theta}$ is modeled on $A_{\theta}$ as follows 
$$
U^{l}_{\bf n} (a) = U_{\bf n} a \, , \qquad (a)U_{\bf n}^{r} = a U_{\bf n}(-1)^{n_{i}} \, , \quad a\in A_{\theta}
$$
 The operator $W^{P}$ is given by   (\ref{w0}). 
The duality transformations $\eta_{i}$ specified by   $E_{\eta_{i}, \theta}$ act on $\tau_{\epsilon}$ (for brevity we 
omit the argument in the notation of these invariants) as follows 
$$
\eta_{1}: \tau_{\epsilon} \mapsto \tau_{ \bar \epsilon_{1} + \bar 1, \epsilon_{2}} \, , \quad 
\eta_{2}: \tau_{\epsilon} \mapsto \tau_{ \epsilon_{1}, \bar \epsilon_{2} + \bar 1} \, . 
$$

 The  bimodules of the second kind correspond to automorphisms of $B_{\theta}$ given by 
$$
U_{\bf n} \mapsto U_{\bf n} e^{\pi i <{\bf n}\theta {\bf k}>} \, , \qquad W \mapsto WU_{\bf k}  \, .
$$ 
The equivalence bimodule is modeled on $A_{\theta}$: 
$$
U^{l}_{\bf n} (a) = U_{\bf n} a \, , \enspace (a)U_{\bf n}^{r} = a U_{\bf n}e^{\pi i <{\bf n}\theta {\bf k}>}  \, , \enspace
W^{P} (a) = w(a)U_{\bf k} \, .
$$
It suffices to take ${\bf k} = (1,0)$ and ${\bf k}=(0,1)$. We denote the corresponding bimodules $E_{\eta^{*}_{i}, \theta}$, $i=1,2$.
The duality transformations $\eta^{*}_{i}$ specified by   $E_{\eta^{*}_{i}, \theta}$ are 
$$
\eta_{i}^{*}:  \tau_{\epsilon} \mapsto (-1)^{\epsilon_{i}} \tau_{\epsilon} \, .
$$

There is one more bimodule $E_{C, \theta}$ generated by 
the automorphism $W\mapsto -W$. Its construction is obvious; in fact it reflects a simple ambiguity in choosing $W^{P}$ 
in any equivalence bimodule. The duality transformation $C$ induced by  $E_{C, \theta}$ reflects the overall sign of the 
invariants $\tau_{\epsilon}$.
It is easy to equip these bimodules with connection $\nabla^{P}$ satisfying the right properties.

One can describe now the full duality group in terms of generators 
$$
\{ R\in SL(2, {\bf Z})_{M} , \, \sigma, \, \tau, \, \eta_{i}, \, \eta_{i}^{*}, \, C \}
$$ 
and relations which are easy to compute from 
(\ref{modular}) - (\ref{sigma}):
\begin{eqnarray*}  
&&R^{-1}\sigma R = \sigma \, , \enspace R^{-1}\tau R \tau = \eta_{1}^{(1+ \bar R_{21} + \bar R_{11})} \eta_{2}^{(1+ \bar R_{22} + \bar R_{12})} \, ,\\
&&\sigma \eta_{1}\sigma  = \eta_{2}^{*} \, , \qquad \sigma \eta_{2} \sigma = \eta_{1}^{*} \, , \\
&& \tau \eta_{i} \tau = \eta_{i} \, , \quad \tau \eta_{1}^{*}\tau = \eta_{1}^{*}\eta_{2} \, , \quad \tau \eta_{2}^{*} \tau = \eta_{2}^{*} \eta_{1} \, ,\\
&& (\eta_{i})^{2} = 1\, , \enspace , (\eta_{i}^{*})^{2} = 1\, , \quad \eta_{i}\eta_{j}^{*}\eta_{i}\eta_{j}^{*}=\delta_{ij}C \, , \\
&& \eta_{i}\eta_{j} = \eta_{j}\eta_{i} \, , \quad \eta^{*}_{i}\eta^{*}_{j} = \eta^{*}_{j}\eta^{*}_{i}\, , \\
&& R^{-1}\eta_{\bar k} R = \eta_{\bar R^{t} \bar k} \, , \quad R^{-1}\eta^{*}_{\bar k} R = \eta^{*}_{\bar R^{-1} \bar k}
\end{eqnarray*}
where $\eta_{\bar k}\equiv \eta_{1}^{\bar k_{1}}\eta_{2}^{\bar k_{2}}$, 
$\eta^{*}_{\bar k}\equiv (\eta^{*}_{1})^{\bar k_{1}}(\eta^{*}_{2})^{\bar k_{2}}$, 
the bar denotes taking ${\rm mod}\, 2$ residues.
To get the complete set of relations one should add to those listed above  the standard relations
 for generators within  $SL(2, {\bf Z})_{M}$ and  $SL(2, {\bf Z})_{T}$.

Using the relations above it is not hard to check that the subgroup generated by 
 $ \tilde \tau = \tau \eta_{1}\eta_{2}C$ and $\sigma$  is isomorphic to $SL(2,{\bf Z})$ and commutes with the 
subgroup $SO(2,2|{\bf Z})_{M}$. Denote this subgroup $SL(2,{\bf Z})_{\tilde T}$.
Then the whole group is isomorphic to a semidirect product 
$ SO(2,2|{\bf Z})\ltimes H$ where $ SO(2,2|{\bf Z})= SL(2, {\bf Z})_{M}\times SL(2,{\bf Z})_{\tilde T}$ and $H$ is a subgroup 
generated by $\eta_{i}$, $\eta_{i}^{*}$ and $C$. The subgroup $H$ is isomorphic   to a central 
extension of $({\bf Z}_{2})^{4}$. The action of the $SO(2,2|{\bf Z})$ subgroup on $H$ follows from the  above commutation 
relations. Reducing entries of matrices from $SO(2,2|{\bf Z})$ modulo 2 we obtain 
a representation of this group on a four-dimensional vector space $({\bf Z_{2}})^{4}$ over the finite field ${\bf Z}_{2}$.
Then the action of $ SO(2,2|{\bf Z})$ on $H$ can be considered as an extension of this representation.


\end{document}